\documentstyle[12pt]{article}
\title{{\normalsize{{\hskip 9cm} BIHEP-TH-95-8, Feb. 1995}}\\
        Electroweak Penguin Effects In Some $B_s^0$ Two-Body
        Decays}

\author{
Dong-sheng Du$^{a,b}$ and Mao-zhi Yang$^b$ \\
        a CCAST(World Laboratory),
        $~~$P.O.Box 8730, Beijing 100080, China\\
        b Institute of High Energy Physics, Academia Sinica,\\
        $~~$P.O.Box 918(4), Beijing, 100039,
        P. R. China\thanks{mailing address.}}
\date{}
\begin{document}
\maketitle
\vspace*{0.3cm}
\begin{center}
\section*{Abstract}
\end{center}

Using the next-to-leading order low energy effective Hamiltonian for
$\Delta B=1$ transitions, the effects of electroweak penguin operators
in some two-body decay modes of $B_s^0$ meson are estimated in the Standard
Model (SM). We find that in $B_s^0\rightarrow\pi^+ K^-$ and
$B_s^0\rightarrow K^+K^-$ decay modes, the electroweak penguin effects are
small, while in $B_s^0\rightarrow\pi^0\bar{K^0}$, $\phi\bar{K^0}$,
$\phi\phi$, the electroweak penguin operators
enhance or reduce the pure QCD penguin and tree level contributions
by $20\%\sim 80\%$ in decay width. We also present the results of CP
asymmetries in these $B_s^0$ deacy modes.
\newpage

Recently Buras et.al. generalized the low energy effective Hamiltonian for
$|\Delta B|=1$ transitions to next-to-leading order QCD corrections[1$\sim$4].
Now, we are in a position to investigate not only the pure QCD penguin and
tree level contributions to B meson decays, but also the effects caused by
electroweak penguin diagrams beyond the leading logarithmic approximation.
Naively, people believe that electroweak penguin contributions to B meson
decays are suppressed by a factor of $\alpha_{em}/\alpha_s\approx{\cal{O}}
(10^{-2})$ relative to QCD penguin contributions [5,6]. So electroweak
corrections comparing with QCD penguin contributions in B system may be
negligible. However this is not always true. In some B meson decay modes,
electroweak penguin contributions can play a significant role[7,8]. In this
paper, we investigate some $B_s^0$ two-body decay channels. They are $B_s^0
\rightarrow\pi^+K^-,~K^+K^-,~\pi^0\bar{K^0},~\phi\phi$, and
$\phi\bar{K^0}$, which are induced by QCD and electroweak penguin diagrams.
We calculate the partial deacy widths, branching ratios, and
the CP asymmetries of these decay modes. We compare the decay
widths including full electroweak penguin contributions with that including
only tree and QCD penguin diagrams. We find that in
$B_s^0\rightarrow \pi^+K^-$, and $K^+K^-$, electroweak correction is small,
it just modify the pure QCD corrections by $1\%\sim 8\%$. While in
$B_s^0\rightarrow \pi^0\bar{K^0},~\phi\bar{K^0}$, and $\phi\phi$,
the electroweak penguin contributions are quite large relative to QCD
penguin corrections. They enhance or reduce the results obtained by only
taking into account tree and QCD penguin operators by $20\%\sim 80\%$.
The decay mode $B_s^0\rightarrow\phi\phi$ has been calculated in [8]. We
include it here just for completeness and comparison. Our results for
this channel agree with that in [8].

In the standard model, the number of colors $N_c$ is 3, while some people
argue that the number favored by experimental data is 2 [9]. So in this paper
we take not only $N_c=3$, but also $N_c=2$. We also present the results
with $N_c=\infty$ as in [10].

The next-to-leading order low energy effective Hamiltonian relevant to
charmless B decays can be taken as the following form [7]:
$$\begin{array}{rl}
H_{eff}(\Delta
B=1)&=~\frac{G_F}{\sqrt{2}}\left[\displaystyle\sum_{q=u,c}v_q\left\{Q_1^q
C_c(\mu)+Q_2^q C_2(\mu)\right.\right.\\[4mm]
&~\left.\left.\displaystyle\sum_{k=3}^{10} Q_k C_k(\mu)\right\}\right],
\end{array}
\eqno(1)$$
\noindent where $C_k(\mu)$ (k=1,$\cdots$,10) are Wilson coefficients which
are calculated in renormalization group improved perturbation theory and
include leading and next-to-leading order QCD corrections .
$v_q$ is the product of Cabibo-Kobayashi-Maskawa (CKM) matrix elements
and defined as
$$
v_q=\left\{
          \begin{array}{ll}
          V_{qd}V_{qb}^*    & \mbox{for $\bar{b}\rightarrow\bar{d} $
                               transitions}\\
          V_{qs}V_{qb}^*    & \mbox{for $\bar{b}\rightarrow\bar{s} $
                               transitions ~~~~~.}
          \end{array}
          \right.
$$
\noindent In our numerical calculations the CKM matrix elements are taken as
$\lambda=0.22$, $A=0.8$, $\eta=0.34$, $\rho=-0.12$ in Wolfenstein
parametrization, which are the preferred values obtained from the fit to the
experimental data[11]. The ten operators are taken as the following form [2,3]
$$\begin{array}{rl}

Q_1^u&=(\bar{b}_{\alpha}u_{\beta})_{V-A}(\bar{u}_{\beta}q_{\alpha})_{V-A},~~~~~~~~~
   Q_2^u=(\bar{b}u)_{V-A}(\bar{u}q)_{V-A},\\
   Q_3&=(\bar{b}q)_{V-A}\displaystyle\sum_{q'}(\bar{q'}q')_{V-A},~~~~~~~~~~~~
   Q_4=(\bar{b}_{\beta}q_{\alpha})_{V-A}
        \displaystyle\sum_{q'}(\bar{q'}_{\alpha}q'_{\beta})_{V-A},   \\
   Q_5&=(\bar{b}q)_{V-A}\displaystyle\sum_{q'}(\bar{q'}q')_{V+A},~~~~~~~~~~
   Q_6=(\bar{b}_{\beta}q_{\alpha})_{V-A}
        \displaystyle\sum_{q'}(\bar{q'}_{\alpha}q'_{\beta})_{V+A},  \\

Q_7&=\frac{3}{2}(\bar{b}q)_{V-A}\displaystyle\sum_{q'}e_{q'}(\bar{q'}q')_{V+A},
   ~~~~~~~~~
   Q_8=\frac{3}{2}(\bar{b}_{\beta}q_{\alpha})_{V-A}
   \displaystyle\sum_{q'}e_{q'}(\bar{q'}_{\alpha}q'_{\beta})_{V+A}, \\

Q_9&=\frac{3}{2}(\bar{b}q)_{V-A}\displaystyle\sum_{q'}e_{q'}(\bar{q'}q')_{V-A},
   ~~~~~~~~~~
   Q_{10}=\frac{3}{2}(\bar{b}_{\beta}q_{\alpha})_{V-A}
        \displaystyle\sum_{q'}e_{q'}(\bar{q'}_{\alpha}q'_{\beta})_{V-A},
 \end{array}
 \eqno(2)$$
\noindent where $Q_1^q$ and $Q_2^q$ are current-current operators, q=u, c,
for q=c case, the two operators $Q_1^c$ and $Q_2^c$ are obtained through
making substitution $u\rightarrow c$ in $Q_1^u$ and $Q_2^u$. $Q_3\sim Q_6$
are QCD penguin operators, the sum $\displaystyle\sum_{q'}$ is runing over all
the quark
flavors being active at $\mu=m_b$ scale, q'=\{u, d, s, c, b\}.
$Q_7\sim Q_{10}$ are electroweak penguin operators, $e_{q'}$ are the electric
charges of the relevant quarks in unit e which is the charge of the proton.
The subscripts $\alpha,~
\beta$ are $SU(3)_c$ color indices. $(V\pm A)$ referes to $\gamma_{\mu}(1\pm
\gamma_5)$.

The Welsion coefficient functions $C_i(\mu)$
are renormalization scheme (RS) dependent beyond the leading order
approximation. Thus we should cancel this renormalization scheme dependence.
So define renormalization scheme independent Wilsion Coefficients[12]
$${\bf \bar{C}}(\mu)=(\hat{1}+\hat{r}_s^T\alpha_s(\mu)/4\pi+
                        \hat{r}_e^T\alpha_{em}(\mu)/4\pi)\cdot
                        {\bf C}(\mu),
                        \eqno(3)$$
\noindent and treat the matrix elements to one-loop level[7],
$$<{\bf Q}^T(\mu)>=<{\bf Q}^T>_0\cdot \left[\hat{1}+\frac{\alpha_s(\mu)}{4\pi}
                      \hat{m}^T_s(\mu)+\frac{\alpha_{em}}{4\pi}\hat{m}^T_e
                      (\mu)\right],\eqno(4)$$
\noindent Combine eq.(3) and (4) we obtain
$$\begin{array}{rl}
  <{\bf Q}^T(\mu)\cdot {\bf C}(\mu)>&=~<{\bf Q}^T>_0\cdot \left[\hat{1}+
  \frac{\alpha_s(\mu)}{4\pi}\left(\hat{m}_s(\mu)-\hat{r}_s\right)^T+
  \frac{\alpha_{em}(\mu)}{4\pi}\left(\hat{m}_e(\mu)-\hat{r}_e\right)^T
  \right]\cdot {\bf \bar{C}}(\mu) \\
  &\equiv ~<{\bf Q}^T>_0\cdot {\bf C'}(\mu)
  \end{array}
  \eqno(5)$$

  The corresponding elements of the matrices $\hat{r}_s$,
  $\hat{r}_e$, $\hat{m}_s(\mu)$ and $\hat{m}_e(\mu)$ are given by
  [1, 4, 13]. Substituting these matrix into eq.(5), we can obtain
  $C'_i(\mu)$ as
  $$\begin{array}{rl}
  C'_1&=~\bar{C}_1,~~~~~~~~~~~~~C'_2~=~\bar{C}_2,\\
  C'_3&=~\bar{C}_3-P_s/N_c,~~~~~~C'_4~=~\bar{C}_4+P_s,\\
  C'_5&=~\bar{C}_5-P_s/N_c,~~~~~~C'_6~=~\bar{C}_6+P_s, \\
  C'_7&=~\bar{C}_7+P_e,~~~~~~~~~C'_8~=~\bar{C}_8,      \\
  C'_9&=~\bar{C}_9+P_e,~~~~~~~~~C'_{10}~=~\bar{C}_{10},
  \end{array} \eqno(6)$$
  \noindent where $P_{s,e}$ are given by
$$\begin{array}{rl}
  P_s&=~\displaystyle\frac{\alpha_s}{8\pi}\bar{C_2}(\mu)
       \left[\frac{10}{9}-G(m_q,q,\mu)\right] , \\[4mm]
  P_e&=~\displaystyle\frac{\alpha_{em}}{3\pi}\left(\bar{C_1}(\mu)+
  \displaystyle\frac{\bar{C_2}(\mu)}{Nc}\right)\left[\frac{10}{9}-
 G(m_q,q,\mu)\right],\\[4mm]
 G(m_q,q,\mu)&=~-4\int_0^1 x(1-x)dxln\displaystyle\frac{m_q^2-x(1-x)q^2}
{\mu^2},
\end{array} $$
 \noindent here q=u,c, for numerical calculation, we take $m_u=0.005GeV$,
$m_c=1.35GeV$, and
use $q^2=m_b^2/2$, which represents the average ``physical'' value.

 We take $m_t=174 GeV$,$m_b=5.0GeV$, $\alpha_s(M_z)=0.118$,
 $\alpha_{em}(M_z)=\frac{1}{128}$,
 and take the numerical values of the renormalization scheme independent
 Wilson coefficients $\bar{C_i}(\mu)$ as [8]
 $$\begin{array}{rl}
 \bar{C_1}&=~-0.313,~~~~\bar{C_2}~=~1.150,~~~~\bar{C_3}~=~0.017,\\
 \bar{C_4}&=~-0.037,~~~~\bar{C_5}~=~0.010,~~~~\bar{C_6}~=~-0.046, \\
 \bar{C_7}&=~-0.001\cdot\alpha_{em},~~~
 \bar{C_8}~=~0.049\cdot\alpha_{em}, ~~~
 \bar{C_9}~=~-1.321\cdot\alpha_{em},   \\
 \bar{C_{10}}&=~0.267\cdot\alpha_{em},
 \end{array}
 \eqno(7)$$

 Using vacuum-saturation approximation the decay amplitude\\
 $<XY|H_{eff}(\Delta B=1)|B_s^0>$ can be factorized into a product of
two current matrix elements $<X|J^{\mu}|0>$ and $<Y|J'_{\mu}|B_s^0>$.
The hadronic matrix elements are calculated in BSW method[14]. We define
 $M^{XY}_{q_1q_2q_3}$ as in Ref.[10].
 $$M^{XY}_{q_1q_2q_3}=\frac{G_F}{\sqrt{2}}<X|(\bar{q_1}q_2)_{V-A}|0>
                       <Y|(\bar{b}q_3)_{V-A}|B>, \eqno(8) $$

 Now we can express the amplitudes $<XY|H_{eff}(\Delta B=1)|B_s^0>$ in
 terms of $C_i'$ and $M^{XY}_{q_1q_2q_3}$.
 $$\begin{array}{rl}
 <\pi^+K^-|H_{eff}|B_s^0>&=~\left[v_u
a_1+\displaystyle\sum_{q=u,c}v_q\left\{a_3
   \right.\right.\\
 &+\displaystyle\frac{M_{\pi}^2}{(m_u+m_d)(m_b-m_u)}(2a_5+3e_u a_7)
  +a_9\cdot\frac{3}{2}e_u
   \left.\right\}\left.\right]\cdot M^{\pi^+K^-}_{udu},\\[4mm]

 <K^+K^-|H_{eff}|B_s^0>&=~\left[v_u a_1+\displaystyle\sum_{q=u,c}v_q\left\{a_3
   \right.\right.\\
 &+\displaystyle\frac{M_{K}^2}{(m_u+m_S)(m_b-m_u)}(2a_5
+3e_u a_7)+a_9\cdot\frac{3}{2}e_u
   \left.\right\}\left.\right]\cdot M^{K^+K^-}_{usu},   \\[4mm]

 <\pi^0\bar{K^0}|H_{eff}|B_s^0>~&=~\left[v_u
a_2+\displaystyle\sum_{q=u,c}v_q\left\{a_4-a_6
    -a_8\cdot \frac{3}{2}e_u+a_{10}\cdot \frac{3}{2}e_u\right\}\right]
    M^{\pi^0\bar{K^0}}_{uud}\\[3mm]
    &+\left[\displaystyle\sum_{q=u,c}\left\{(\frac{1}{N_c}+1)(C'_3+C'_4)-a_6-
    a_8\cdot\frac{3}{2}e_d+\displaystyle\frac{M_{\pi^0}^2}{m_d(m_b-m_d)}
    (a_5+\displaystyle\frac{3}{2}e_d\cdot a_7)\right.\right.\\
    &\left.\left.+(\frac{1}{N_c}+1)\cdot\frac{3}{2}e_d(C'_9+C'_{10})
    \right\}\right]M^{\pi^0\bar{K^0}}_{ddd},\\[4mm]

<\phi\phi|H_{eff}|B_s^0>&=~\left[\displaystyle\sum_{q=u,c}v_q\left\{(1+\frac{1}{N_c})
     (C'_3+C'_4)+a_6+\frac{3}{2}e_s a_8 \right.\right.\\
     &\left.\left.+(1+\frac{1}{N_c})(C'_9+C'_{10})\cdot
     e_s\right\}\right]M^{\phi\phi}_{sss},\\[4mm]

 <\phi\bar{K^0}|H_{eff}|B_s^0>&=~\left[\displaystyle\sum_{q=u,c}v_q\left\{a_3+
 \frac{M^2_{k^0}}{(m_s+m_d)(m_s+m_b)}(-2a_5-3e_sa_7)+\frac{3}{2}e_sa_9
 \right\}\right]M^{\bar{K^0}\phi}_{sds}\\[3mm]
     &+\left[\displaystyle\sum_{q=u,c}\left\{a_4+a_6+
     \frac{3}{2}e_s(a_8+a_{10})\right\}\right]
     M^{\phi\bar{K^0}}_{ssd},
\end{array} \eqno(9)$$
\noindent where the masses of $d$ quark and $s$ quark are taken as
$m_d=0.01GeV$, $m_s=0.175GeV$ in numerical calculation, and $a_k$ is defined as
$$\begin{array}{rl}
a_{2i-1}&\equiv~\displaystyle\frac{C'_{2i-1}}{N_c}+C'_{2i},\\
a_{2i}&\equiv~C'_{2i-1}+\displaystyle\frac{C'_{2i}}{N_c},~~~(i=1,2,3,4,5)
\end{array} $$

The authers of Ref.[8] have calculated
$<\phi\phi|H_{eff}|\bar{B_s^0}>$. Our results in eq.(9) are
consistent with theirs after making a charge conjugation. We follow
the definition of relevant decay constants and form factors in Ref.[14],
and use eq.(9$\sim$ 13) in Ref.[10] to calculate
$\displaystyle\sum_{q=u,c}|M^{XY}_{q_1q_2q_3}|^2$. In the rest frame
of $B_s^0$, the two body
decay width is calculated by
$$\Gamma(B_s^0\rightarrow XY)=\frac{1}{8\pi}|<XY|H_{eff}|B_s^0>|^2\frac{|p|}
{M^2_B}, \eqno(10)$$
\noindent where $|p|=\frac{[(M_B^2-(M_X+M_Y)^2)(M_B^2-(M_X-M_Y)^2)]^{
                     \frac{1}{2}}}{2M_B}$
 is the magnitude of momentum of the particles X and Y. $M_B$, $M_X$ and
 $M_Y$ are the mass of $B_s^0$, X and Y, respectively. For calculating
the branching ratio, the total decay width of $B_s^0$ is taken as
$\Gamma_{tot}=4.91\times 10^{-13} GeV$[15]. We give our
 numerical results in Table 1-5, where``QCD+EW'' means the decay width or
 branching ratio with full QCD penguin and EW penguin corrections, ``QCD''
with only QCD penguin corrections. $\frac{\Gamma_{QCD+EW}-\Gamma_{QCD}}
 {\Gamma_{QCD}}$ represents the enhancement percentage of EW penguin
 contributions to the decay width.
 $R\left(\frac{QCD+EW}{tree}\right) $, $R\left(\frac{EW}{tree}\right)$,
 and $R\left(\frac{QCD}{tree}\right)$ represent the relevant ratioes of
 amplitudes of tree and penguins.

{}From Table 1 and 2, we can see that for $B_s^0\rightarrow \pi^+K^-$ and
$K^+K^-$, the EW penguin effects are small, they only reduce or enhance
the decay width by $0.6\%\sim 8\%$, and the results are not sensitive to $N_c$.
For $B_s^0\rightarrow \pi^+K^-$, the tree level contribution is dominant,
while for $B_s^0\rightarrow K^+K^-$, the QCD penguin contribution dominant.
{}From Table 3 - 5, we find that for $B_s^0\rightarrow \pi^0\bar{K^0}$,
$\phi\bar{K^0}$
and $\phi\phi$, the EW penguin correction is significant. The
percentage of correction is almost as high as $-80.3\%$ for $B_s^0\rightarrow
\phi\bar{K^0}$ decay modes when taking $N_c=3$.

In Table 6 and 7, we present the branching ratios and CP asymmetry
parameters of the five $B_s^0$ decay modes discussed above. For $f$ being
non-CP-eigenstate cases, the CP asymmetry  is defined as the following
$$
  {\displaystyle\cal{A}}_{cp}=\frac{\Gamma(B_s^0\rightarrow f)-
              \Gamma(\bar{B_s^0}\rightarrow\bar{f})
        }{\Gamma(B_s^0\rightarrow f)+\Gamma(\bar{B_s^0}\rightarrow\bar{f})
        }, \eqno(11)
$$
while for $f$ being CP-eigenstate case, we calculate the CP asymmetry
through[16]
  $$ \begin{array}{rl}
  {\displaystyle\cal{A}}_{cp}&=~\displaystyle\frac{\int_0^{\infty}
           [\Gamma(B_{phys}^0 (t)\rightarrow f)-
              \Gamma(\bar B_{phys}^0 (t)\rightarrow f)]dt
             }{\int_0^{\infty}
           [\Gamma(B_{phys}^0 (t)\rightarrow f)+
              \Gamma(\bar B_{phys}^0 (t)\rightarrow f)]dt}\\[4mm]
            &=~\displaystyle\frac{1-|\xi|^2-2Im\xi(\Delta m/\Gamma)}
      {(1+|\xi|^2)[1+(\Delta m/\Gamma)^2]},
\end{array} \eqno(12)
$$
\noindent where $\Delta m/\Gamma\simeq 20$ which is the preferred value
in the Standard Model[11].

We find that only for $B_s^0\rightarrow \pi^0\bar{K^0}$, and $\phi\bar{K^0}$
two cases, CP violation parameter is sensitive to the number of colors $N_c$,
the other four decay modes are not.

\vspace*{0.2cm}
This work is supported in part by the China National Natural Science
Foundation and the Grant of State Commision of Science and Technology of
China.

\newpage
\begin{center}
Table 1 Numerical results of $B_s^0\rightarrow\pi^+K^-$
\end{center}
\begin{center}
\begin{tabular}{|c|c|c|c|c|c|c|}\hline
      & \multicolumn{2}{c|}{$\Gamma$ ($10^{-18}GeV$)}&
$\frac{\Gamma_{QCD+EW}-\Gamma_{QCD}}{\Gamma_{QCD}}$&
    R($\frac{QCD+EW}{tree}$)&R($\frac{QCD}{tree}$)
    &R($\frac{EW}{tree}$)\\ \cline{2-3}
      & QCD+EW & QCD &      &     &     &     \\ \hline
 $N_c=2$&1.83  &1.86  & -1.23\% &0.197 & 0.187 & 0.010\\ \hline
 $N_c=3$&2.016 & 2.025& -0.42\% &0.197 & 0.193 & 0.004 \\ \hline
 $N_c=\infty$&2.45  & 2.42 &0.69\%&0.194 &0.200 &0.006 \\ \hline
\end{tabular}
\end{center}

\begin{center}
Table 2 Numerical results of $B_s^0\rightarrow K^+K^-$
\end{center}
\begin{center}
\begin{tabular}{|c|c|c|c|c|c|c|}\hline
      & \multicolumn{2}{c|}{$\Gamma$ ($10^{-18}GeV$)}&
$\frac{\Gamma_{QCD+EW}-\Gamma_{QCD}}{\Gamma_{QCD}}$ &
    R($\frac{QCD+EW}{tree}$)&R($\frac{QCD}{tree}$)
    &R($\frac{EW}{tree}$)\\ \cline{2-3}
      & QCD+EW & QCD &      &     &     &     \\ \hline
 $N_c=2$&3.19   & 2.93& 8.61\% &3.61 & 3.43& 0.178\\ \hline
 $N_c=3$&3.58   & 3.48& 2.93\% &3.62 & 3.56& 0.065\\ \hline
 $N_c=\infty$&4.22&4.43 &-4.82\%&3.56& 3.68& 0.114 \\ \hline
\end{tabular}
\end{center}
\begin{center}
Table 3 Numerical results of $B_s^0\rightarrow\pi^0\bar{K^0}$
\end{center}
\begin{center}
\begin{tabular}{|c|c|c|c|c|c|c|}\hline
      & \multicolumn{2}{c|}{$\Gamma$ ($10^{-19}GeV$)}&
$\frac{\Gamma_{QCD+EW}-\Gamma_{QCD}}{\Gamma_{QCD}}$&
    R($\frac{QCD+EW}{tree}$)&R($\frac{QCD}{tree}$)
    &R($\frac{EW}{tree}$)\\ \cline{2-3}
      & QCD+EW & QCD &      &     &     &     \\ \hline
 $N_c=2$&1.59  &1.97  & -19.0\% &0.426 & 0.617 & 0.195\\ \hline
 $N_c=3$&0.45  &0.69  & -34.7\% &1.831 & 2.516 & 0.703 \\ \hline
 $N_c=\infty$& 0.59   & 0.58  & 1.72\% & 0.501 & 0.647 &0.149\\ \hline
\end{tabular}
\end{center}
\begin{center}
Table 4 Numerical results of $B_s^0\rightarrow\phi \bar{K^0}$
\end{center}
\begin{center}
\begin{tabular}{|c|c|c|c|}\hline
      & \multicolumn{2}{c|}{$\Gamma$ ($10^{-20}GeV$)}&
$\frac{\Gamma_{QCD+EW}-\Gamma_{QCD}}{\Gamma_{QCD}}$\\ \cline{1-3}
      & QCD+EW & QCD &           \\ \hline
 $N_c=2$&1.78 &3.46 & -48.6\% \\ \hline
 $N_c=3$&0.12 &0.62 & -80.3\%  \\ \hline
 $N_c=\infty$&3.50 &2.22 &58.0\% \\ \hline
\end{tabular}
\end{center}

\begin{center}
Table 5 Numerical results of $B_s^0\rightarrow\phi \phi$
\end{center}
\begin{center}
\begin{tabular}{|c|c|c|c|}\hline
      & \multicolumn{2}{c|}{$\Gamma$ ($10^{-18}GeV$)}&
$\frac{\Gamma_{QCD+EW}-\Gamma_{QCD}}{\Gamma_{QCD}}$\\ \cline{1-3}
      & QCD+EW & QCD &           \\ \hline
 $N_c=2$&2.46 &3.27 & -24.8\% \\ \hline
 $N_c=3$&1.56 &2.11 & -26.1\%  \\ \hline
 $N_c=\infty$&0.38 &0.52 &-26.9\% \\ \hline
\end{tabular}
\end{center}

\begin{center}
Table 6   Branching ratioes and CP asymmetry parameters \\
          with only tree and QCD penguin contributions
\end{center}
\begin{center}
\begin{tabular}{|c|c|c|c|c|c|c|}\hline
decay mode& \multicolumn{3}{c|}{Br}&
            \multicolumn{3}{c|}{{\cal{A}}$_{cp}$} \\ \cline{2-7}
          &$N_c=2$ &$N_c=3$& $N_c=\infty$
          &$N_c=2$ &$N_c=3$& $N_c=\infty$ \\ \hline
$B_s^0\rightarrow \pi^+K^-$& $3.78\times  10^{-6}$&$4.12\times 10^{-6}$&
  $4.95\times 10^{-6}$ &-7.0\%& -7.9\% &-8.2\% \\ \hline
$B_s^0\rightarrow K^+K^-$& $5.97\times 10^{-6}$& $7.08\times 10^{-6}$&
  $9.02\times 10^{-6}$& -2.31\%&-2.24\%&-2.18\% \\ \hline
$B_s^0\rightarrow \pi^0\bar{K^0}$&$ 4.00\times10^{-7}$&$1.40\times10^{-7}$&
  $1.18\times10^{-7}$&8.7\%&6.3\%&-32.6\%\\ \hline
$B_s^0\rightarrow \phi\bar{K^0}$&$7.06\times10^{-8}$&$1.25\times10^{-8}$&
  $4.52\times10^{-8}$&-1.35\%&-5.0\%&0.97\% \\ \hline
$B_s^0\rightarrow \phi\phi$&$6.67\times10^{-6}$&$4.29\times10^{-6}$&
  $1.06\times10^{-6}$&-0.010\%&-0.014\%&-0.024\%         \\ \hline
\end{tabular}
\end{center}
\begin{center}
Table 7   Branching ratioes and CP asymmetry parameters \\
          with full tree , QCD and EW penguin contributions
\end{center}
\begin{center}
\begin{tabular}{|c|c|c|c|c|c|c|}\hline
decay mode& \multicolumn{3}{c|}{Br}&
            \multicolumn{3}{c|}{{\cal{A}}$_{cp}$} \\ \cline{2-7}
          &$N_c=2$ &$N_c=3$& $N_c=\infty$
          &$N_c=2$ &$N_c=3$& $N_c=\infty$ \\ \hline
$B_s^0\rightarrow \pi^+K^-$& $3.73\times  10^{-6}$&$4.11\times 10^{-6}$&
  $4.98\times 10^{-6}$ & -7.28\%& -8.01\% &-8.12\% \\ \hline
$B_s^0\rightarrow K^+K^-$& $6.49\times 10^{-6}$& $7.29\times 10^{-6}$&
  $8.59\times 10^{-6}$& -2.22\%&-2.21\%&-2.24\% \\ \hline
$B_s^0\rightarrow \pi^0\bar{K^0}$&$ 3.24\times10^{-7}$&$9.15\times10^{-8}$&
  $1.20\times10^{-7}$&10.9\%&10.5\%&-32.0\%\\ \hline
$B_s^0\rightarrow \phi\bar{K^0}$&$3.62\times10^{-8}$&$2.48\times10^{-9}$&
  $7.13\times10^{-8}$&-1.58\%&-13.8\%&0.95\% \\ \hline
$B_s^0\rightarrow \phi\phi$&$5.02\times10^{-6}$&$3.19\times10^{-6}$&
  $7.66\times10^{-7}$&-0.011\%&-0.016\%&-0.023\%\\ \hline
\end{tabular}
\end{center}

\end{document}